# Dielectric Properties and Ion Transport in Layered MoS$_2$ Grown by Vapor-Phase Sulfurization

M. Belete[1,2], S. Kataria[1], U. Koch[3], M. Kruth[4,5], C. Engelhard[3], J. Mayer[4,5],

O. Engström[2], M. C. Lemme[1,2]

[1]RWTH Aachen University, Faculty of Electrical Engineering and Information Technology, Chair of Electronic Devices, Otto-Blumenthal-Str. 2, 52074 Aachen, Germany

[2]AMO GmbH, Advanced Microelectronic Center Aachen, Otto-Blumenthal-Str. 25, 52074 Aachen, Germany

[3]University of Siegen, Department of Chemistry and Biology, Adolf-Reichwein Str. 2, 57076 Siegen, Germany

[4]RWTH Aachen University, Central Facility for Electron Microscopy, Ahornstr. 55, 52074 Aachen, Germany

[5]Ernst Ruska-Centre for Microscopy and Spectroscopy with Electrons, Research Centre Jülich, 52425 Jülich, Germany

*E-Mail:* max.lemme@*eld.*rwth-aachen.de

**Abstract:**

Electronic and dielectric properties of vapor-phase grown $MoS_2$ have been investigated in metal/$MoS_2$/silicon capacitor structures by capacitance-voltage and conductance-voltage techniques. Analytical methods confirm the $MoS_2$ layered structure, the presence of interfacial silicon oxide ($SiO_x$) and the composition of the films. Electrical characteristics in combination with theoretical considerations quantify the concentration of electron states at the interface between Si and a 2.5 - 3 nm thick silicon oxide interlayer between Si and $MoS_2$. Measurements under electric field stress indicate the existence of mobile ions in $MoS_2$ that interact with interface states. Based on time-of-flight secondary ion mass spectrometry, we propose $OH^-$ ions as probable candidates responsible for the observations. The dielectric constant of the vapor-phase grown $MoS_2$ extracted from CV measurements at 100 KHz is 2.6 – 2.9. The present study advances the understanding of defects and interface states in $MoS_2$. It also indicates opportunities for ion-based plasticity in 2D material devices for neuromorphic computing applications.

**Introduction:**

Transition metal dichalcogenides (TMDs) are among a large family of two-dimensional (2D) layered materials. They are generically described by the formula $MX_2$, where M represents a transition metal such as molybdenum (Mo), tungsten (W), niobium (Nb), and others and X stands for a chalcogenide element, i.e. sulfur (S), selenium (Se) or tellurium (Te). [1,2] Bulk TMDs are formed from vertically stacked 2D-layers that are held together by van der Waals forces, typically at an interlayer spacing of less than 1 nm. The electronic properties of TMDs range from semiconducting to superconducting, and include direct and indirect energy band gaps that depend on the number of layers.[1] Molybdenum disulfide ($MoS_2$) is a semiconducting TMD material with a band-gap ranging from 1.3 eV in bulk to 1.88 eV as a mono-layer.[1,3] Its appealing properties have made it a potential material for applications in nanoelectronics, optoelectronics and neuromorphic computing.[4–13] Most of the early-stage research has been conducted on small flakes obtained through mechanical exfoliation[14] or chemical exfoliation.[15,16] Although these techniques can yield high quality $MoS_2$, they are not suitable for large scale applications as they are limited to small flakes.[17] More recently, CVD growth from gaseous precursors and thermal conversion of metal films (vapor-phase sulfurization) have been proposed as scalable methods for $MoS_2$ growth.[18–22] Here, the latter technique has been used to grow large area $MoS_2$ directly on silicon (Si) substrates. Most of the research on electronic devices based on $MoS_2$ has focused on lateral transport properties of $MoS_2$. In this work, we investigate layered $MoS_2$ on Si substrates with respect to transport perpendicular to the interfaces, for a potential application as a dielectric barrier material in vertical heterostructure devices, such as graphene and 2D materials-based hot electron transistors (HETs) which are potential devices for high speed electronics.[23–27] The small band gap of $MoS_2$ compared to available oxides,[1,3] and its low band offset with respect to silicon (Si)[1,3,28] makes it a good candidate for efficient emission barriers in HETs.[29] In combination with an ultra-thin dielectric layer, $MoS_2$ could also serve in a bi-layer tunnel barrier configuration, which has been shown to enhance HET on-current levels.[30] In this context, it is essential to understand the electronic and dielectric properties of $MoS_2$ as a barrier material. We have thus investigated capacitors, with $MoS_2$ as the dielectric, through capacitance-voltage (C-V) and conductance-voltage (G-V) measurements. Raman spectroscopy ($\lambda$ = 532 nm) and transmission electron microscopy (TEM) were carried out to obtain information on the phase formation and structure of the $MoS_2$ films, respectively. The chemical composition of the samples was investigated with time-of-flight secondary ion mass spectrometry (ToF-SIMS) depth profiling.

Metal-Semiconductor-Semiconductor (MSS) capacitors in which $MoS_2$ is sandwiched between a metal electrode and a Si substrate were fabricated on p- and n-type Si substrates. The $MoS_2$ was converted from ~ 5 nm molybdenum (Mo) thin films deposited on the samples. In the conversion process, the metal films were sulfurized by heating them in a furnace at 800 °C in argon and sulfur atmosphere, resulting in ~ 15 nm thick $MoS_2$ films. Afterwards, circular metal electrodes with diameters of 100 µm were formed through a sequence of photolithography, deposition of a stack of chromium (Cr, 20 nm) and gold (Au, 120 nm) by thermal evaporation, and a lift-off process. Finally, the native oxide on the backside of the samples was removed by HF and a stack of Cr and Au was deposited as a back side contact. A schematic of the process flow is presented in Fig. 1a. The setup used for the vapor-phase sulfurization process is illustrated in Fig. 1b and an optical micrograph of a device is shown in Fig. 1c.

The MoS$_2$ films were characterized by Raman spectroscopy after growth. The Raman spectra show the two prominent peaks ($E^1_{2g}$ and $A_{1g}$, Fig. 2a), indicating the formation of the crystalline MoS$_2$ phase.[31,32] The $E^1_{2g}$ peak is attributed to the in-plane vibrations of Mo and S atoms, while the $A_{1g}$ peak signifies the out-of-plane vibrations of S atoms.[32] Cross-sectional TEM investigations have been employed to visualize the layer sequence pertained in the MSS structure and the associated interfaces. Fig. 2b reveals a clearly layered structure of the MoS$_2$ film, which is nanocrystalline in nature and where most of the layers are oriented nearly vertical with respect to the Si face. Such vertically aligned layers are commonly observed in thick MoS$_2$ and other TMD films grown by vapor-phase sulfurization technique.[33,34] The TEM images further show a ~ 2.5 to 3 nm thick amorphous SiO$_X$ interfacial layer (IL) between Si and MoS$_2$, which is consistent with earlier reports involving MoS$_2$ growth on Si.[22,33] The presence and formation of this SiO$_x$ layer can be attributed to: (1) post-HF treatment re-growth of native oxide on the Si surface before the Mo film deposition, and (2) oxidation of the Si surface by oxygen from water molecules intercalated at the Si-Mo interface while heating up the samples during the sulfurization process. ToF-SIMS measurements provide a depth profile of the chemical composition of the samples (Fig. 2c). In addition to the expected elemental / molecular composition of the intended layers, we observed the presence of negatively charged chloride (Cl$^-$) and hydroxyl (OH$^-$) ions in the MSS structures. See methods section for details on the ToF-SIMS measurements.

**Results and Discussion:**

C-V and G-V measurements were carried out inside a Lakeshore cryogenic probe station connected to a Keithley KI-590 admittance meter in vacuum ($10^{-4}$ mbar) and at room temperature. C-V characteristics measured at 100 KHz signal frequency on the MoS$_2$ capacitors with p- and n- Si substrates are shown in Fig. 3. The shape of these graphs resembles that of typical metal/oxide/silicon (MOS) structures,[35] with capacitance saturation over a large voltage range for the p-type samples (Fig. 3a). The n-type samples also exhibit saturation (Fig. 3b), but leakage current dominates for gate voltages above 4 V. As a consequence, the measured capacitance starts to drop below the saturation level above that point (not shown). Based on the C-V measurements, we calculated the dielectric constant values for the vapor-phase grown MoS$_2$ using

$$C_{ins.} = \frac{C_{SiO_x} C_{MoS_2}}{C_{SiO_x} + C_{MoS_2}}, \qquad (1)$$

where the insulator capacitance $C_{ins.}$ is analogous to the oxide capacitance of conventional MOS capacitors, $C_{SiO_x}$ is the IL capacitance and $C_{MoS_2}$ is the MoS$_2$ capacitance.

In this calculation, $C_{ins.}$ is considered to be the equivalent capacitance of the SiO$_x$ and the MoS$_2$ capacitors connected in series, and it is obtained from the saturation part of the C-V curves shown in Fig. 3. The extracted dielectric constant values are in the range of 2.6 – 2.9. Santos and Kaxiras [36] suggested that the dielectric constant of MoS$_2$ varies with applied external electric fields (E$_{ext}$) and number of layers. In our measurements, the electric fields at which the capacitance started to saturate were in the range of 0.0033 V/Å – 0.02 V/Å. For this field range, the predicted MoS$_2$ dielectric constant of approximately 3 [36] is in reasonable agreement with the extracted values of the present work. However, we note that the theory was based on horizontally aligned MoS$_2$ layers, while the experiments were carried out on vertically aligned ones (Fig. 2b).

The C-V measurements indicate that energy barriers exist between Si and MoS$_2$ for both holes and electrons, and that they are sufficiently high to establish accumulation of carriers at the Si band edges (i.e. saturating CV curves). The magnitude of these barriers depends on energy band alignment, which could vary between the following two extremes:

(1):- Assuming that the band alignment is entirely determined by electron affinities, one would expect a "Type - I" (straddling) gap,[37] where the barrier heights are determined by the difference between the electron affinities of the two materials in contact.

(2):-Assuming that the alignment is mainly controlled by the existence of "virtual gap states", a "Type - II" (staggered) gap [37] will form. In this case, the hole barrier is expected to be considerably larger than the electron barrier, which is in accordance with the present observation of current leakage mentioned earlier. In reality, however, one may expect a combination of the two phenomena.[38] The presence of two interfaces (i.e. Si/SiO$_x$ and SiO$_x$/MoS$_2$) in the present devices makes the situation even more complicated as the band alignment can be influenced by virtual gap states in both SiO$_x$ and MoS$_2$ layers. Based on the saturation effects observed in the C-V data discussed above, we propose simplified energy band schemes (Figs. 3c and 3d) for the present MSS structures. These band schemes feature a SiO$_x$ transition layer, where $x \leq 2$, between Si and MoS$_2$. Compared to a standard pure SiO$_2$ gate oxide, this interfacial layer seems to be more permeable to charge carriers. The leakage current through the $SiO_x$ transition layer in the present MSS structures (Fig. S3) was examined and found to be about three orders of magnitude higher than leakage current reported for standard $SiO_2$ of comparable thickness and applied gate voltage.[39] This indicates that the interfacial layer in the current device is of poor quality, possibly due to oxygen deficiencies and other defects favoring leakage.

For further investigation, we performed bias-stress (BS) measurements on the MSS devices. BS measurements allow investigating charge dynamics in dielectrics and at their interfaces, and also to study the effect of interface states on the C-V and G-V characteristics. Here, a reference C-V curve was first measured without BS with 70 ms delay between data points. Next, bias stress of $V_{Gstress}$ = 4 V and -4 V was applied for 1 min, followed by C-V and G-V measurements that were run with the shortest possible delay between data points (i.e.1 ms). This resulted in shifting of the curves along the voltage axis. Subsequent BS cycles were applied until the curves did not exhibit further observable shifts. Fig. 4a shows C-V characteristics for the MSS capacitors on n-Si measured under positive BS of $V_{Gstress}$ = 4 V, which shifts the curves to the left along the voltage axis. In addition, the curves exhibit a hump that increases for increasing positive BS cycles. A similar development can also be observed in the G-V plots presented in Fig 4b. Here, the growing maxima in the G-V curves is a clear indication of the existence of a peak in the $D_{it}$ distribution. A second maximum develops that quickly becomes larger in value than the initial maximum of the G-V curves. The horizontal shift and increase in the hump intensity of the C-V curves are consistent with the observed shift and increase in the peak intensity of the G-V curves. These results indicate influence by interface states.[40,41] Negative BS measurements (with $V_{Gstress}$ = -4 V) were conducted on the same device. This time, the C-V and G-V curves exhibit parallel shifts in the opposite direction as for positive BS, i.e. to the right along the voltage axis (Figs 4c and 4d,). In contrast to positive BS, humps were not observed on the negative BS C-V curves and the maximum of the G-V peaks showed a gentle decreasing trend for increasing negative BS cycle number. In addition, the voltage shift was smaller than for positive BS. Similar trends were observed in the C-V and G-V characteristics of samples with p-Si substrates under BS (Figs. 5a and 5b), albeit with an additional "turnaround" effect after the first negative BS cycle, which will be discussed in detail later (Figs. 5c and 5d). Similar trends were observed for both positive and negative BS C-V and G-V measurements carried out on other devices (see Supporting Information, Figs. S1 and S2).

The C-V data was further analyzed through simulations based on an equivalent circuit model.[41] This was inspired by the experimental evidence of negatively charged ions within $MoS_2$ that may move in response to BS and interact with interface states to influence the capacitance measurements. The circuit configuration given by the capacitance meter to yield the measured quantities is depicted in Fig. 6a. This configuration contains a capacitor and a conductor connected in parallel. However, the physical MSS system should be represented by more circuit elements (Fig. 6b). This equivalent circuit model takes into account the capacitance contributions from the IL, $C_{SiO_x}$, $C_{MoS_2}$ and $C_s$ (silicon depletion capacitance), and also considers capacitance and conductance contributions from the interface states, $C_{it}$ and $G_{it}$, respectively. The capacitance and conductance contributions from interface states are initiated by the oscillation of the Si Fermi-level position (Δμ) with respect to the energy position (ΔE) of the interface states with an energy distribution, $D_{it}$ (ΔE). Depending on the energy position of the interface states with respect to the Fermi-level, charge carriers are captured and emitted by the states in pace with the angular frequency, ω, of the measurement AC signal. This periodic transfer of charge carriers from and into the interface states establishes the interface state capacitance, $C_{it}$, and a conductance, $G_{it}$, which influence the final measured quantities, $C_m$ and $G_m$. In addition, since the Fermi-function is not a precise step function at finite temperatures, the capacitance meter cannot distinguish between capacitance contributions from charge carriers captured by traps with energy levels at the exact position of the Fermi-level and those within the tails of the Fermi function. To address this issue, it is appropriate to introduce a concept of capacitance density ($\chi_{it}$), which represents the capacitance per unit area and unit energy.[35,41] The mathematical expression for this quantity can be formulated as

$$\chi_{it} = \frac{q^2 D_{it} 2e_n^2}{2k_B T \left(4e_n^2 + \omega^2\right)} f(1-f), \qquad (2)$$

where q is the electron charge, $D_{it}$ is the energy distribution of interface states, $e_n$ is the rate of emission of electrons at the trap, $k_B$ is Boltzmann's constant, T is the absolute temperature, and $f$ is the Fermi-function.

Integrating the capacitance density along ΔE results in the interface state capacitance, $C_{it}$, for a given Fermi-level position (Δμ):

$$C_{it} = \int_0^{E_g} \chi_{it} d(\Delta E), \qquad (3)$$

The electron emission rates at the interface states are assumed to be much higher than the frequency of the AC probe signal from the capacitance meter, so that the trap states are completely emptied and filled within a period time of the AC signal. Therefore, the measured differential capacitance $C_m$ can be calculated as [35,41]

$$C_m = \frac{C_{MoS_2} C_{SiO_x}(C_s+C_{it})}{C_{MoS_2} C_{SiO_x} + (C_s+C_{it})(C_{MoS_2}+ C_{SiO_x})}, \quad (4)$$

where, $C_m$ is the measured capacitance, $C_{MoS2}$ is the MoS$_2$ capacitance, $C_{SiO_x}$ is the IL (SiO$_x$) capacitance, $C_s$ stands for the silicon capacitance and $C_{it}$ is the interface state capacitance as calculated using Eqn. 3.

The theoretical capacitance was calculated using Eq.4 and compared with experimental data. Fig. 6c shows measured C-V curves (symbols) at four BS cycles and the corresponding theoretical fits (solid curves) by using $D_{it}$ distributions shown in Fig. 6d. Each $D_{it}$ distribution was individually tuned to fit the respective theoretical C-V curve with the corresponding measured data. The model accurately describes the experimental data, including the amplitudes of the observed C-V humps (Fig 6c). These results confirm that the increasing amplitude of the C-V humps originate from the peaks in the $D_{it}$ distribution of interface states. The model in combination with the good fit to the experimental data also suggests that the active interface states, $D_{it}$, changes in concentration and energy maximum as a function of positive BS cycles.

These horizontal and vertical shifts of the C-V and G-V curves in response to BS can be attributed to the movement of negative mobile charges inside the MoS$_2$ bulk, and their interactions with interface electron states. Under negative BS, mobile negative charges are pushed towards the IL-MoS$_2$ interface, leading to positive parallel shifts in the C-V and G-V curves as observed in Figs. 4c and 4d. This situation is similar to the electric field induced movement of sodium ions in SiO$_2$ studied in the early days of MOS development.[42] Furthermore, as the negative mobile charges approach the interface, coulomb forces may disturb the electron potential of the states. This would in turn influence their energy positions and even give rise to a passivation effect similar to that caused by hydrogen in SiO$_2$,[43] leading to the observed reduction of humps in the C-V curves (Fig. 4c and 4d). On the other hand, positive BS drags mobile negative charges away from the IL-MoS$_2$ interface, thus removing their influence on the interface states. This causes a decrease in the negative charge at the interface and an increase in concentration of active interface states that can capture and emit electrons. As a result, negative voltage shifts and pronounced humps are observed in the C-V curves in Fig. 4a. The same effect was also manifested in the negative shifts and increasing peaks of the G-V curves (Fig. 4b). Such effect indicates increasing interface state concentration. The shape of the interface state distributions resemble that of the P$_b$ centers occurring at SiO$_x$/Si interfaces.[39,40] Their presence is attributed to the interfacial SiO$_x$ layer formed on Si, which is very similar to the well-documented examples from research towards integration of high-k oxides in Si MOSFETs.[40,41]

Finally, the "turn-around" effect observed in the C – V and G-V curves for p-type samples after the first negative BS cycle warrants a discussion. Here, an initial shift of the curves in the negative voltage direction is followed by positive voltage shifts for succeeding stress cycles (Figs. 5c and 5d). This indicates that either an increasing positive charge or a decreasing negative charge occurs close to the silicon side of the capacitor after the first cycle, but not after those following. Taking into consideration the existence of hole accumulation at the Si/SiO$_x$ interface during negative BS, conceivable origins of this effect could be the interaction of holes either with traps of the SiO$_x$ layer or the moving ions, thus leaving a more positive oxide charge behind. Based on the ToF-SIMS data, we propose that negatively charged mobile ions are responsible for the observed BS responses. In particular, due to a high possibility for adsorption of water molecules in the MoS$_2$ layer during and after device fabrication, we consider hydroxyl ions ($OH^-$) as probable candidates. These would be expected to originate from water splitting due to the catalytic properties of Cr and MoS$_2$ in aqueous solutions.[46–50] It should also be noted that the random and predominantly vertical orientation of the MoS$_2$ layers in our devices may facilitate the mobility of the $OH^-$ ions along the van der Waals gaps which are oriented parallel to the electric field.

**Conclusions:**

Capacitors with vapor phase-grown MoS$_2$ layers as dielectric were fabricated and characterized in detail. TEM images revealed a nanocrystalline nature of MoS$_2$ and the formation of an amorphous SiO$_x$ interfacial layer between Si and MoS$_2$. The extracted dielectric constants of MoS$_2$ are in the range of 2.6 – 2.9 for electric fields ≤ 0.02 V/Å. In addition, ToF-SIMS depth profiles suggest the presence of OH$^-$ ions in the devices. From theoretical calculations that were fitted to the experimental data, P$_b$-like peaks and their concentrations indicate the presence of significant amount of interface states and confirm the presence of mobile negative ions. Such features have so far been largely neglected in experiments using layered $MoS_2$ as an electronic material, as is the case in few-layer $MoS_2$ FETs.

This study may serve as an indication of challenges to come as the research on transition metal chalcogenides moves from exfoliated (near-perfect) flakes to materials grown with scalable, semiconductor fabrication compatible methods. Ion- and defect-based [8] phenomena in 2D materials may also be exploited in future neuromorphic computers, where neuroplasticity is a required feature.

**Methods:**

Metal-Semiconductor-Semiconductor (MSS) capacitors in which $MoS_2$ is sandwiched between a metal and Si were fabricated on p- and n-type silicon (Si) substrates. First, hydrofluoric acid (HF) was used to remove native silicon oxide from the Si surface, followed by an e-beam evaporation of ~ 5 nm molybdenum (Mo) thin film on the samples. Then, the samples were heated in a furnace at 800 °C in argon and sulfur atmosphere, resulting in ~ 15 nm thick $MoS_2$ films. Afterwards, circular metal gates with a diameter of 100 μm were formed through a sequence of photolithography, deposition of a stack of chromium (Cr, 20 nm) and gold (Au, 120 nm) by thermal evaporation, followed by a lift-off process. Finally, the native oxide on the back side of the samples was removed by HF and a stack of Cr and Au was deposited. A schematic diagram showing the complete process flow for fabricating the MSS capacitors is presented in Fig. 1a. The vapor-phase sulfurization process is illustrated in Fig. 1b and an optical microscope top view image of the as-fabricated device is shown in Fig. 1c.

ToF-SIMS measurements began by sputtering a target area on the samples with oxygen ($O_2$) ions at 5 KeV, which resulted in a 500 x 500 μm$^2$ crater. This first step was carried out to remove possible contaminations on the top-most layers of the samples and reduce errors in the measurements. Then, bismuth ions ($Bi^+$) were used at 25 keV to carry out the actual depth profile measurements within a smaller area (200 x 200 μm$^2$) at the center of the larger crater. Each individual measurement was carried out after a 5 s pre-sputtering step. Due to the fact that the sample was sputtered with oxygen, metals form oxides which appear in the mass spectra at relatively high signal abundance. The profiles presented in this work were measured in the negative ion mode, as the species of interest were best accessible that way.

**Acknowledgements:**

The authors are indebted to Gregor Schulte (University of Siegen) for his kind help in the deposition of Molybdenum films and Jasper Ruhkopf (RWTH Aachen University) and Andreas Bablich (University of Siegen) for their assistance with the C-V measurement setup. Support from the Micro- and Nanoanalytics Facility at the University of Siegen is gratefully acknowledged. Financial support from the European Commission through the European Regional Development Fund (HEA2D, 0801002) and the Graphene Flagship (785219) is gratefully acknowledged.

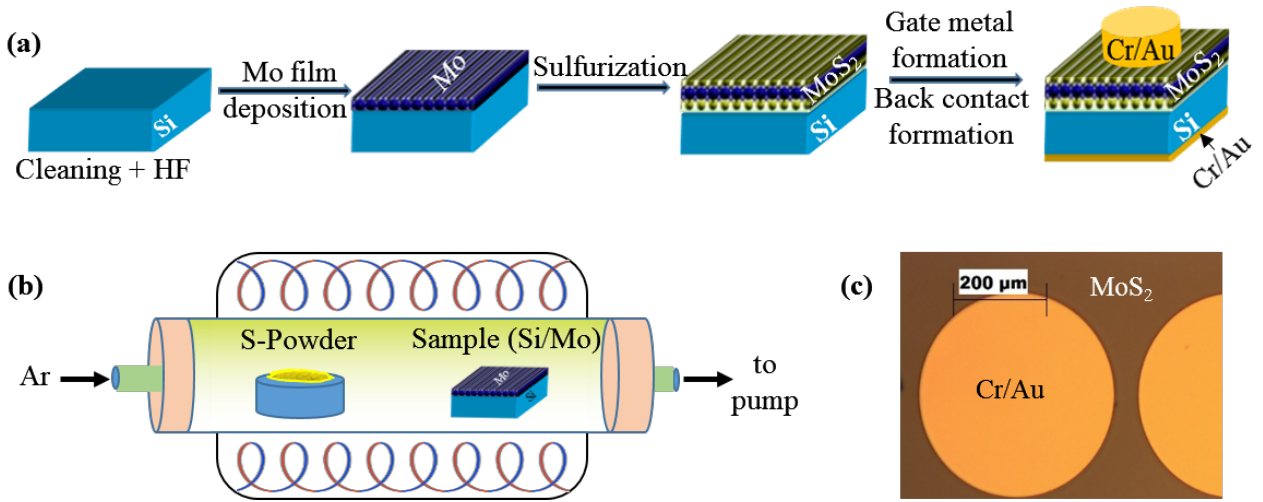

Figure 1: (a) Schematic diagram illustrating the fabrication process flow for MSS capacitors, (b) Schematic diagram showing the vapor-phase sulfurization growth process for the MoS$_2$ films used as a dielectric in the MSS capacitors. (c) Top-view optical microscope image of the fabricated MSS capacitor.

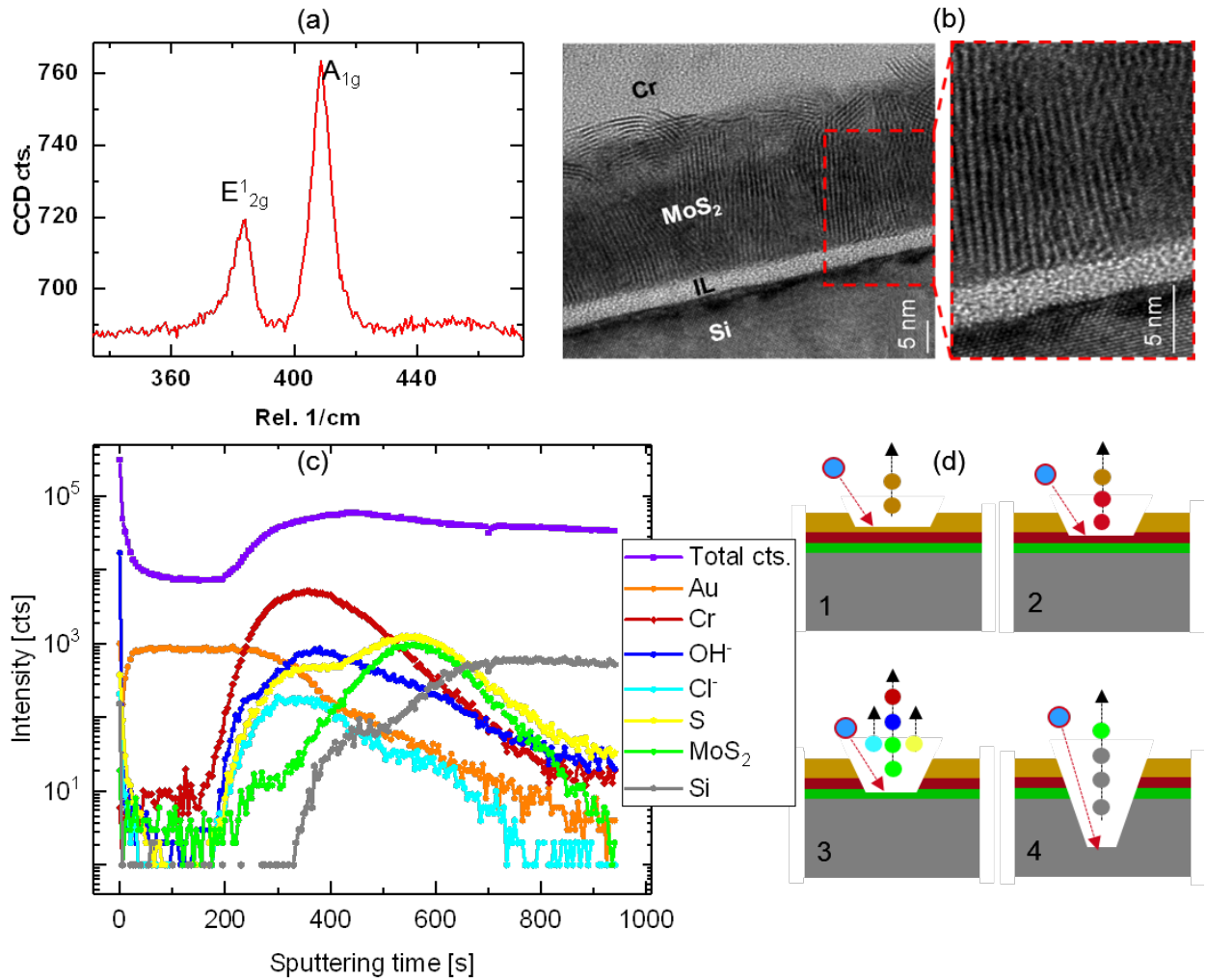

Figure 2: (a) Raman spectrum of a vapor phase grown $MoS_2$ film. (b) Cross-sectional transmission electron micrograph of an MSS structure indicating that the majority of the nanocrystalline $MoS_2$ layers is aligned vertically. The formation of an $SiO_x$ interfacial layer is also visible at the $Si$-$MoS_2$ interface. (c) ToF-SIMS depth profile measurements confirming the presence of $OH^-$ ions in the MSS structures. The SIMS profile further confirms the expected elemental composition of the MSS structure, i.e. silicon, sulfur, $MoS_2$, chromium and gold. Profiles of identical species originating from each layer were added to obtain a good representation of the composition of the layers in the structure. To access the species of interest, the profiles were measured in the negative ion mode,

which normally puts a negative loading on the species. (d) Schematic diagram illustrating the ToF-SIMS depth profile measurements, in which a heavier incident ion knocks out species from the target sample. Here the numbering 1 - 4 indicates the measurement sequence and the solid circles represent the species presented in the depth profile in (c).

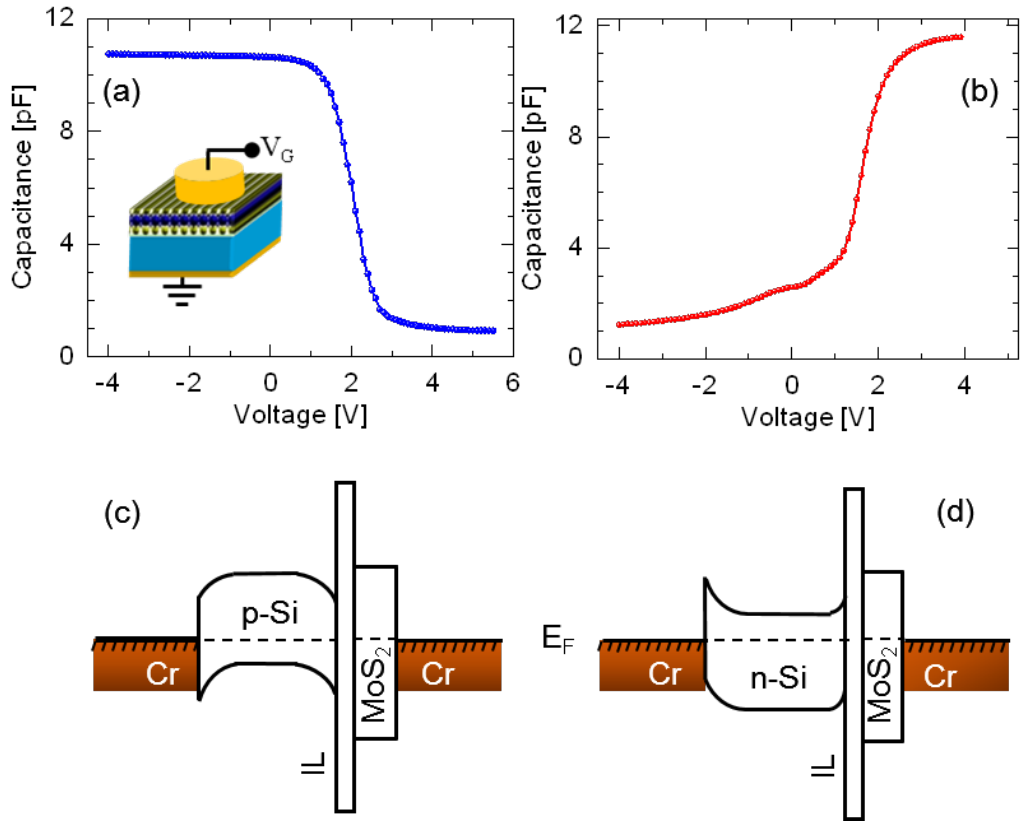

Figure 3: C-V measurements on MSS capacitors with (a) p-type Si and (b) n-type Si at 100 KHz AC signal frequency. Inset (a): Schematic of the device structure with the wiring setup used during measurements. Schematics of band diagrams of the MSS capacitors with (c) p-type, and (d) n-type Si, at thermal equilibrium.

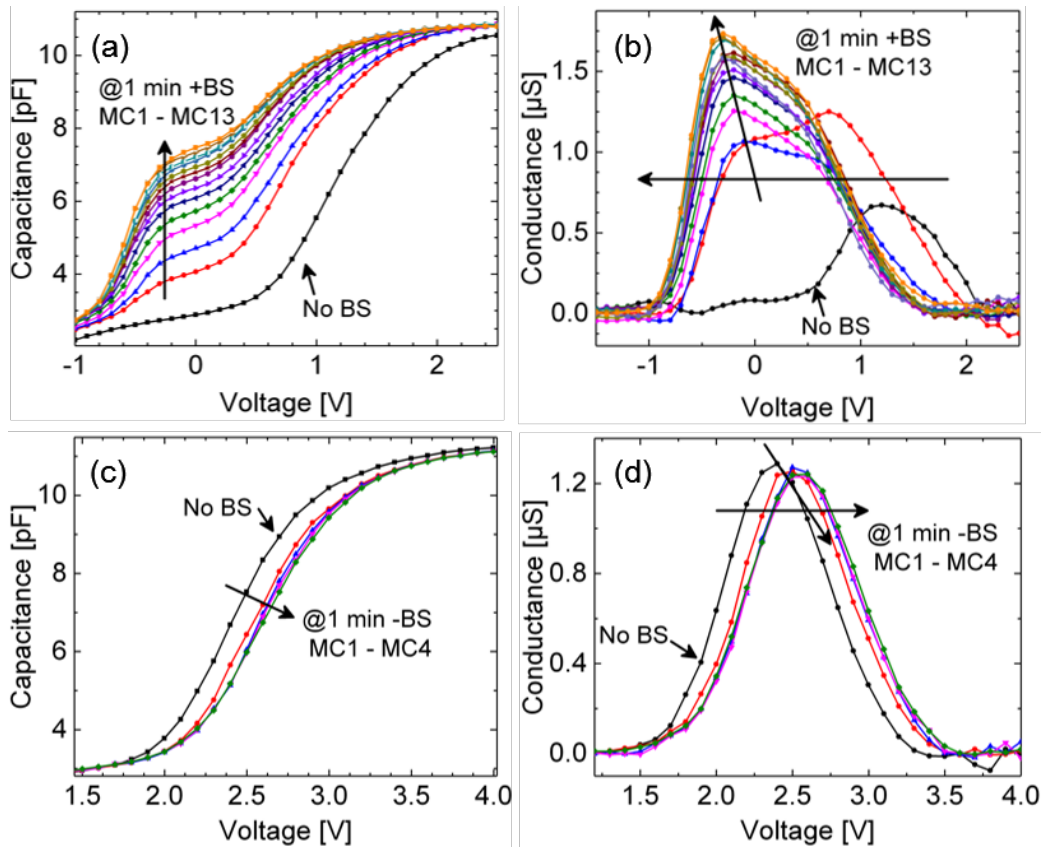

Figure 4: BS C-V and G-V measurements on MSS capacitors with n-Si at 100 KHz AC signal frequency: (a) C-V characteristics showing clear shifts of the capacitance curves to the left under positive BS. (b) Corresponding G-V plots confirming the shifting trend. The voltages at which the conductance maxima occurs correspond to bias voltages assigned to distinct humps in the C-V curves. (c) C-V measurements under negative BS, where the capacitance curves shift to the right. (d) Corresponding G-V plots with a similar shift direction and with peaks of slightly decreasing amplitude. In all measurements, the first black curves were measured without BS and with a 70 ms delay between data points. The remaining curves were measured after 1 min BS and with 1 ms delay. However, for clarity reasons, the graphs in Figs. 4 and 5 present magnified versions of the measurements in a smaller voltage range. In addition,

negative BS measurements were done first and positive BS next, but again for conveniency during discussions they are presented in the opposite order.

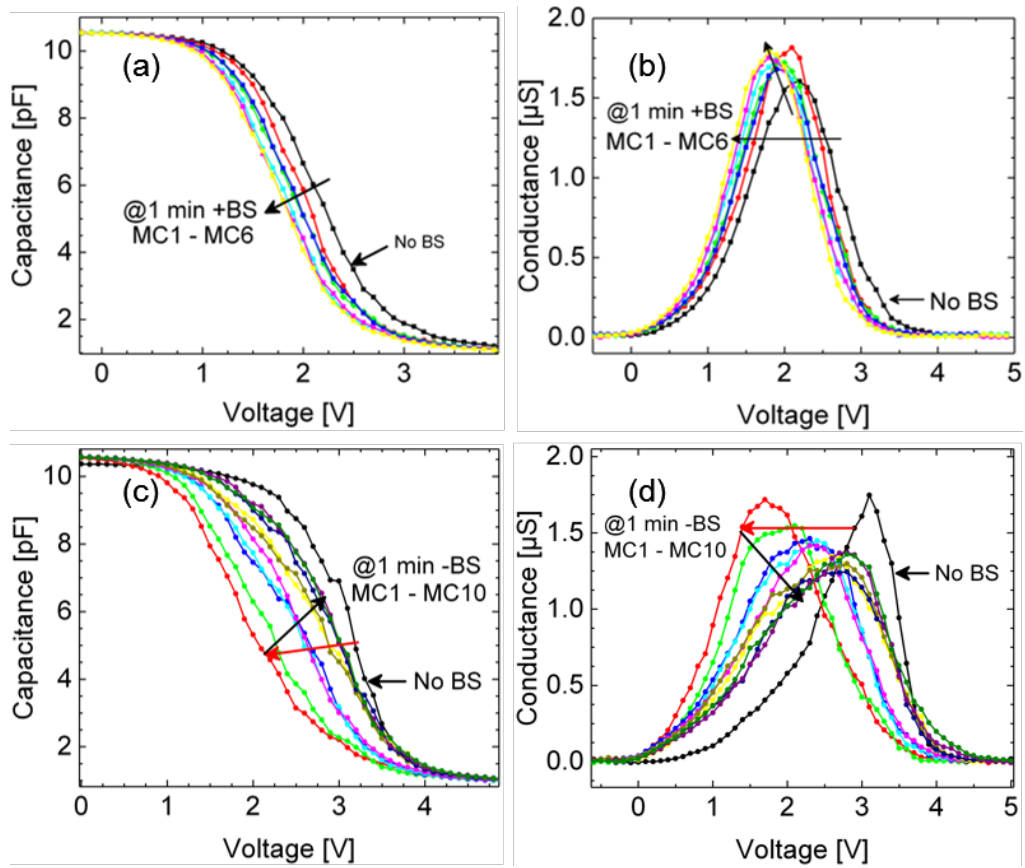

Figure *5*: BS C-V and G-V measurements on MSS capacitors with p-Si at 100 KHz AC signal frequency: (a) C-V characteristics with clear shifts of the capacitance curves to the left under positive BS. (b) Corresponding G-V plots with peaks showing slight increase in amplitude and shifting to the left. (c) C-V measurements on the same device under negative BS. A "turn around" effect is observed: the first BS leads to a shift towards the left, but upon the second bias-stress cycle, the curves shift to the right. (d) Corresponding G-V plots with similar behavior. In all these measurements, the first black curves were measured without BS with a 70 ms delay between data points. The subsequent curves were measured after 1 min BS and with 1 ms delay.

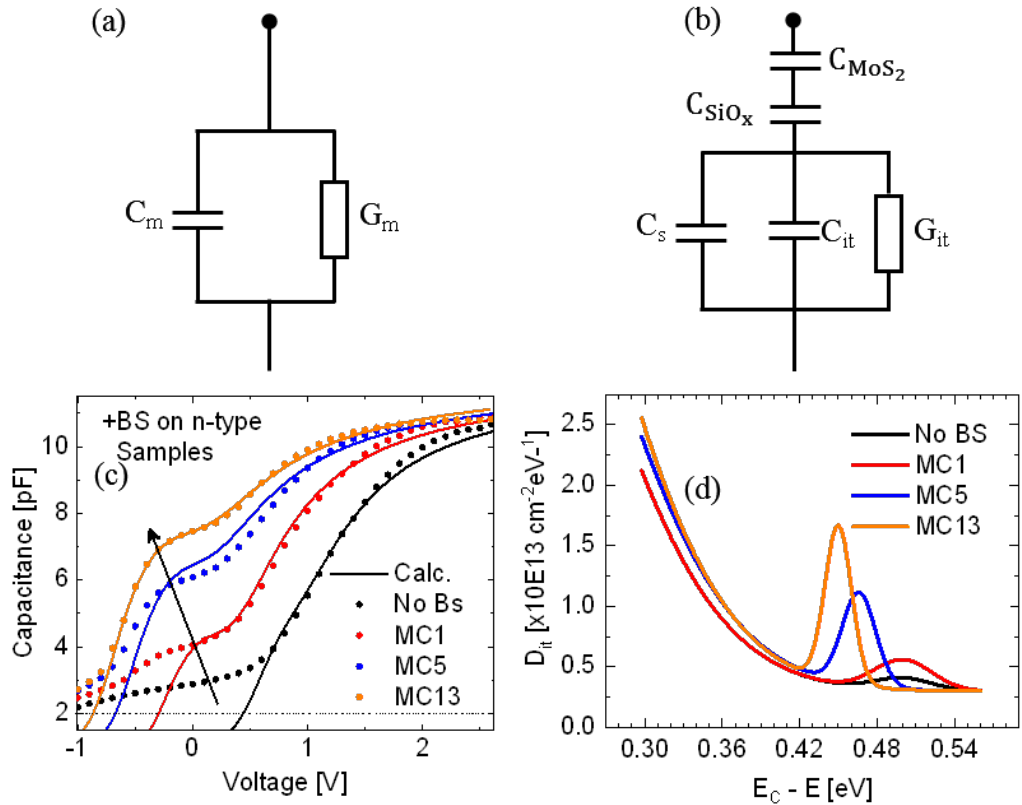

Figure 6: Calculation of C-V curves and fitting them to the experimental C-V data from MSS capacitors on n-Si measured under positive BS: (a) Equivalent circuit model with elements as measured by the C-V meter. (b) Equivalent circuit model with circuit elements representing the physical SSM device under test. (c) Calculated C-V curves (solid lines) fitted to the experimental C-V data measured at 100 KHz (dots). The first curve from the right represents data measured without BS and MC1, MC5 and MC13 are C-V curves at the 1st, 5th and 13th BS cycle, respectively. (d) $D_{it}$ distributions assumed to fit the theoretical C-V curves to the experimental data. Good agreement between the simulated and measured C-V data in combination with the $D_{it}$ trend strongly suggests that the negative C-V shift and the growing hump under positive BS is a result of the mobile negative ions in $MoS_2$ moving away from the $SiO_x$-$MoS_2$ interface, thereby reducing the electrostatic passivation of interface states.

# Supplementary Information

# Dielectric Properties and Ion Transport in Layered MoS$_2$ Grown by Vapor-Phase Sulfurization

M Belete[1,2], S. Kataria[1], U. Koch[3], C. Engelhard[3], M. Kruth[4,5], J. Mayer[4,5],

O. Engström[2], M. C. Lemme[1,2]

[1]RWTH Aachen University, Faculty of Electrical Engineering and Information Technology, Chair of Electronic Devices, Otto-Blumenthal-Str. 2, 52074 Aachen, Germany

[2]AMO GmbH, Advanced Microelectronic Center Aachen, Otto-Blumenthal-Str. 25, 52074 Aachen, Germany

[3]University of Siegen, Department of Chemistry and Biology, Adolf-Reichwein Str. 2, 57076 Siegen, Germany

[4]RWTH Aachen University, Central Facility for Electron Microscopy, Ahornstr. 55, 52074 Aachen, Germany

[5]Ernst Ruska-Centre for Microscopy and Spectroscopy with Electrons, Research Centre Jülich, 52425 Jülich, Germany

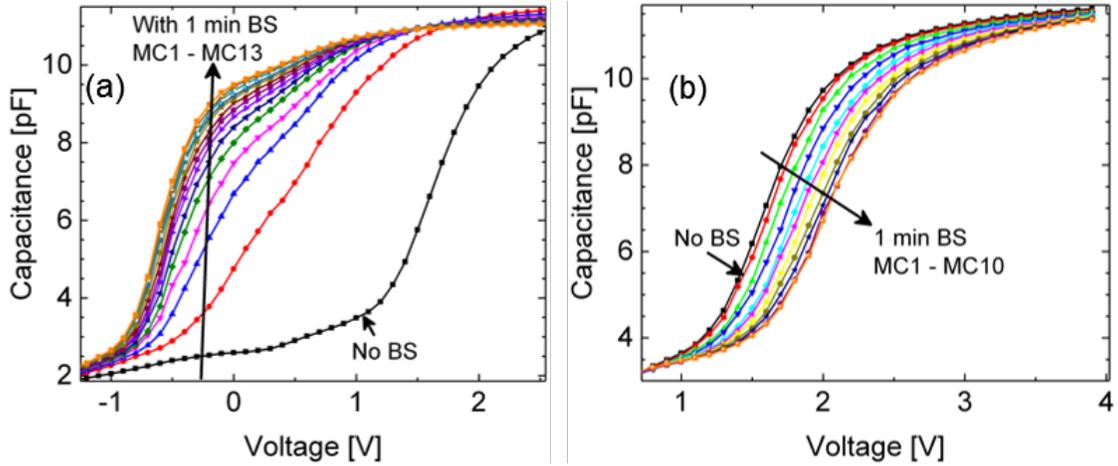

Figure. S1: C-V measurements of MSS capacitors on n-type Si measured under (a) positive bias-stress and (b) negative bias-stress. These measurements were done on a device different from the one reported in Fig. 4 in the main manuscript. However, both measurements exhibit similar trends.

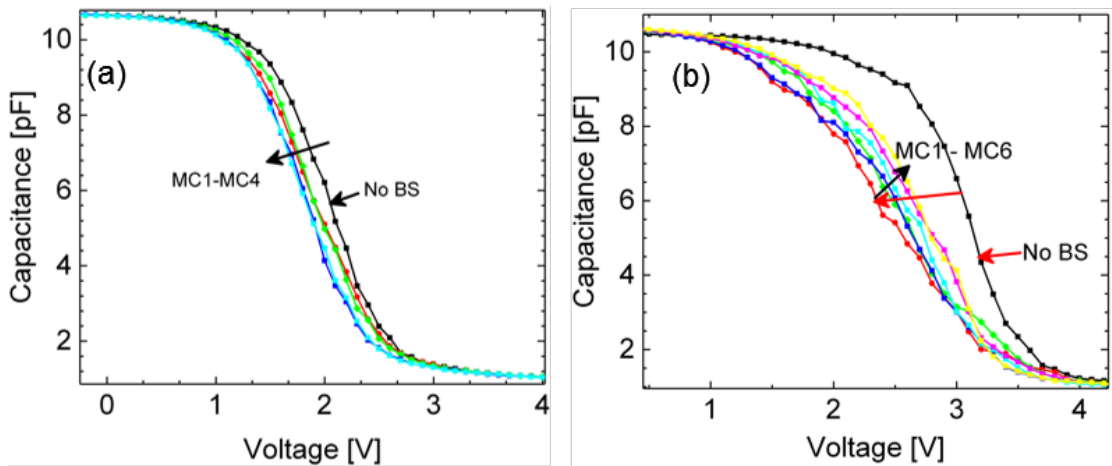

Figure S2: C-V measurements of MSS capacitors on p-type Si, measured under (a) positive bias-stress and (b) negative bias-stress. These measurements were done on a device different from the one reported in Fig. 5 in the main manuscript. Nevertheless,

both measurements show similar trends, including the "turn-around" effect discussed in the main manuscript.

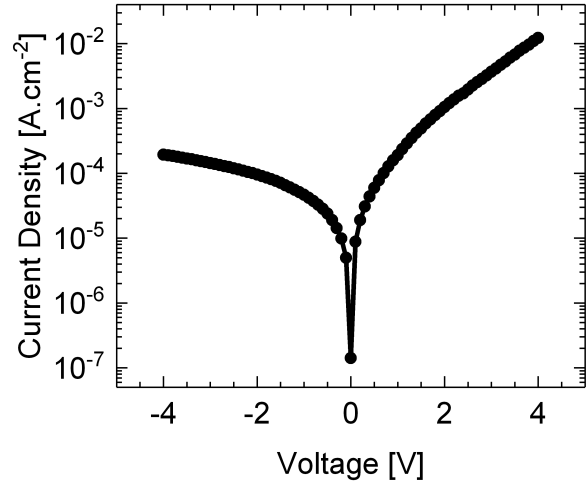

Figure S3: Current Voltage (I-V) characteristics of an MSS capacitor on n-type Si. The data shows clear asymmetry between the electron and hole current branches, indicating different band alignments of the two junctions. The current level in this graph at a given gate voltage is larger by at least three orders of magnitude compared to levels reported for a high-quality $SiO_2$ gate oxides of similar thickness [39].